\documentclass[preprint]{aastex}
\begin{document}
\title{ THE PRECESSION OF MERCURY AND THE DEFLECTION 
OF STARLIGHT FROM SPECIAL RELATIVITY ALONE}
\author{Robert L. Kurucz}
\affil{Harvard-Smithsonian Center for Astrophysics}
\affil{60 Garden St, Cambridge, MA 02138, USA}

\email{rkurucz@cfa.harvard.edu}

\begin{abstract}

I show that the precession of the orbit of Mercury and the deflection of 
starlight by the Sun are effects of special relativity alone when the
gravitational field of a particle is treated in the same way as the
electric field of a charged particle.  General 
relativity is not needed to explain them. [Submitted to Astrophysical 
Journal Letters on 29 June 2006; revised 30 January 2007]

\end{abstract}

\section{The Precession of the Perihelion, or The Deflection of a ``Massy" Particle}

Predicting the precession of the orbit of Mercury is a standard test for 
relativity theories.  The observed precession is 13.489"/orbit, mostly from
precession of the 
observer's coordinate system and from interactions with other bodies in the
solar system (cf. Misner, Thorne, \& Wheeler 1973, p.1113).  Only 
0.104"$\pm$0.002"/orbit of the precession is produced by volume effects
for Mercury and for the Sun and by non-radial relativistic effects (Shapiro et al. 1972).  
But since we do not yet know the internal structure of Mercury (or the Sun), the 
precession of Mercury is not a test of relativity.  In fact the density 
distribution in Mercury can be estimated by requiring that it produce most
of the observed precession.

     Assume that the gravitational field produced by a massy particle   
varies with velocity and has the same properties as the electric field produced 
by a moving electric point charge that is described in electricity and magnetism 
textbooks.  From Resnick (1968) substituting GM for q/4$\pi \epsilon_{0}$,
\begin{equation}
\vec{f} =  (1-\beta^{2})/[1-\beta^{2} \sin^{2} \alpha]^{3/2} \ GM\hat{r}/r^{2},
\end{equation}
where $\beta = v/c, v$ is the particle velocity, c is the 
speed of light, $\alpha$ is the angle between $\hat{v}$ and
$\hat{r}$, G is the gravitational constant,  
M = M$_{0}$/(1-$\beta^{2})^{1/2}$ is the gravitational mass, and M$_0$ is the rest mass.  
The field moves away from the poles, the direction of motion, toward the
equator as the velocity increases.  The $\alpha\beta$ factor is $(1-\beta^2)$
at the poles which goes to 0 as $v$ approaches c.  The $\alpha\beta$ factor 
is $1/(1-\beta^2)^{1/2}$ at the equator which becomes large as $v$ approaches c.
Substituting 
sin$^{2}\alpha$ = 1--cos$^{2}\alpha$ = 1--$(\hat{v}\cdot\hat{r})^{2}$,
\begin{equation}
\vec{f} =  (1-\beta^{2})/[1-\beta^{2}(1-(\hat{v}\cdot\hat{r})^{2}]^{3/2} \ GM\hat{r}/r^{2}. 
\end{equation}

The integral of the force over the volumes of Mercury and the Sun does not 
degenerate into a two-body problem but it can be treated as a two-body problem in 
heliocentric coordinates with perturbations.  The subscripts {\small S} and {\small M}
refer to the Sun and Mercury.  
\begin{equation}
\vec{F} =  - G M_M (M_{\odot}+M_M)(1-v^2/c^2)/[1-v^2/c^2(1-(\hat{v}\cdot\hat{r})^{2}]^{3/2} \ \hat{r}/r^{2}.
\end{equation}
where $\hat{r}$ and $\hat{v}$ are the position and velocity of the center of mass 
of Mercury relative to the position and velocity of the center of mass 
of the Sun, which are defined to be 0.  The two-body force is purely radial.  The 
acceleration toward the Sun is $\vec{a} = \vec{F}/M_M$.  From the calculations 
described below, typical values in this problem are:
period  88 days; r = 0.387 AU or 58 million km; $v$ = 48 km/s; $\beta$ = 0.00016;
$1-\beta^2$ = 0.999999974; M$_{0M}$ = 3.3$\times$10$^{26}$ g; M$_{0M}$/M$_{\odot}$ = 0.000000166;
M$_M$ = 1.0000000128$\times$3.3$\times$10$^{26}$ g; $\hat{v}\cdot\hat{r}$ = -0.20 to +0.20. 
The precession is about 29 km/orbit.

The perturbative forces are defined at the same time and positions as the 
two-body force, at the center of mass of the Sun and the center of mass of 
Mercury.  Define $\vec{r}_S$ and $\vec{v}_S$ as the position and velocity 
vectors of a mass element in the Sun relative to the center of the Sun 
and $\vec{r}_M$ and  $\vec{v}_M$ are the position and velocity 
vectors of a mass element in Mercury relative to the center of mass of Mercury.
Let $\vec{d} = \vec{r} + \vec{r}_S + \vec{r}_M$ and
$\vec{w} = \vec{v} + \vec{v}_S + \vec{v}_M$.
Mercury and the Sun are far apart so they present small solid 
angles to each other.  Assume that the Sun is spherical with density that 
decreases radially from 148 to 0 g/cm$^3$ (Lebretton \& Dappen 1988).  Assume 
that the Sun rotates 
with a surface equitorial velocity of about 2 km/s and that internal motions 
are smaller than 2 km/s and symmetric about the equator.  In the solar part of 
the integrand of the force, angular effects and retardation effects are small
(which I have tested by numerical integration as in Section 2) so that  $\vec{r}_S$ and 
$\vec{v}_S$ can be ignored and the solar part of the integrand can be factored 
out.  Assume that Mercury is spherical with density that decreases radially  
from about 9 to about 3 g/cm$^3$ (Schubert et al. 1988).  A mostly iron 
core fills about three-fourths of the 2440 km radius and a rocky mantle is 
the outer fourth.  The rotation of Mercury is small and internal motions are 
small or non-existent so $\vec{v}_M$ is small and  $\vec{w} = \vec{v}$. 
The force reduces to
\begin{equation}
\vec{F}=-GM_M(M_{\odot}+M_M)\int_M (1-v^2/c^2)/[1-(v^2/c^2)(1-(\hat{v}\cdot\hat{d})^{2}]^{3/2} \rho_M \ \hat{d}/d^2 dV_M/M_M,
\end{equation}
where $\vec{d} = \vec{r} + \vec{r}_M$ and where 
all the mass points are retarded to the center of mass of Mercury.  
The retardation in time is $dt = -(d-r)/c$, and in position is $\vec{d}\ ' = \vec{d}-\vec{v} (d-r)/c$.
This force has non-radial components.
Since $\beta$ is small the denominator can be expanded to yield
\begin{equation}
\vec{F} = -GM_M(M_{\odot}+M_M)\int_M [\small 1+{\small 1/2}\beta^{2}-{\small 3/2}
\beta^{2}(\hat{v}\cdot\hat{d})^{2}]\ \rho_M \ \hat{d}/d^{2} \ dV_M/M_M.
\end{equation}

I approximately computed the precession as follows:  First the orbit was computed 
as a two-body problem with [3] using Butcher's 5th order Runge-Kutta method following 
Boulet (1991) for 50000 steps/day.  The perihelion was determined by 6-point
Lagrangian differentiation.  The two-body orbit repeated the 
perihelion to 100 $\mu$arcsec.  A vectorial correction factor to the two-body 
force for the volume of Mercury was tabulated five times per day for that orbit by 
integrating [5] over 500 density shells, 500 latitudes, and 1000 longitudes 
for a range of Mercury models that tabulate density as a function of radius.
The precession for a uniform density of 
5.426 g/cm${^3}$ was 0.117"/orbit.  An iron-core-with-thin-mantle model that 
I made up inspired by Figure 1 in Schubert et al. (1988) yielded a precession of 
0.107"/orbit.  Thus, using only special relativity, a simple model reproduces 
the precession to within 3 percent. Keeping the thickness of the mantle about 
one fourth the radius,  I tried ad hoc fixes to the density variation
in the model until a precession of 0.105"/orbit was achieved in agreement 
with observation.  Several different models produced the same result.  Remember
that other small volume effects in Mercury and the Sun were ignored in this
calculation; neither is actually spherical.  Seismic 
measurements on Mercury itself will eventually allow a real model to 
be determined that will test the relativistic calculation.

\section{The Deflection of Starlight, or The Deflection of a Massless Particle}

The gravitational field expands from a particle at the speed of light.  
But a massless particle moves at the speed of light.  In the direction
of motion there can be no gravitational field.  In analogy with the relativistic
contraction of the field of a massy particle described above, the whole 
gravitational field of a massless particle is in the plane perpendicular 
to the direction of motion.  Instead of filling a solid angle 4$\pi$, the
field contracts to the ``width" of the equator $d\phi$ with solid angle 
2$\pi d\phi$ and ``strength" $S$.  Then $S2\pi d\phi$ = 4$\pi$ and $Sd\phi$ = 2$\delta_{\perp}$
where $\delta_{\perp}$ is a $\delta$-function in the equatorial plane. The gravitational 
mass of a photon, or other massless particle is 2$\delta$$_{\perp}$E/c$^{2}$.
The gravitational mass averaged over all directions is $E/c^{2}$ and the momentum
of a massless particle is $(E/c)\hat{c}$.

A massless particle gains energy, blueshifts, as it approaches a massive
body B and loses energy, redshifts, as it departs.  The gain in energy is also 
a gain in mass.  The gravitational mass $M_{E}$ is $E/c^{2} = (1+GM_B/r/c^2)E_{r=\infty}$.
The two-body force on a massless particle passing a massive body 
B in B-centric coordinates is 
\begin{equation}
\vec{F} = -GM_BM_E \ 2 \sin \alpha \ \hat{p}/r^2
= -GM_BM_E \  2 \sqrt{1-(\hat{c}\cdot\hat{r})^2} \ \hat{p}/r^2,
\end{equation}
where $\hat{r}$ is the vector from the center of mass of B to the massless 
particle, $\alpha$ is the angle between $\hat{r}$ and $\hat{c}$, and 
$\hat{p}$ is perpendicular to $\vec{c}$, $\hat{c}\cdot\hat{p} = 0$, and lies 
in the plane defined by $\vec{c}$ and $\vec{r}$.  The acceleration is $\vec{a} = \vec{F}/M_E$.
The general case for an extended body B with internal motion is 
\begin{equation}
\vec{F} =  -GM_E \ 2\int_B \rho \sqrt{1-(\hat{c}\cdot\hat{d})^2} \ \hat{p}/d^2 \ dV,
\end{equation}
where $\vec{d} = \vec{r} + \vec{r}_B$ and the mass points are retarded 
to the center of mass of B, $\hat{r}_B$  and $\vec{v}_B$ 
are the position and velocity vectors of a mass element in B relative to the 
position and velocity of the center of mass of B which are defined to be 0.
Here $\hat{p}$ is perpendicular to $\vec{c}$, $\hat{c}\cdot\hat{p} = 0$, 
and lies in the plane defined by $\vec{c}$ and $\vec{d}$.  The retardation in 
time is $dt = -(d-r)/c$, and in position is $\vec{d}\ ' = \vec{d}-\vec{v}_B (d-r)/c$.

     There are no observations of deflection by point masses.  The observations 
are of lensing by ill-defined bodies far away and lensing by the 
Sun.  The solar observations are that photons from a distant star or planet are 
deflected by 1.751$\pm$0.002 $R_{\odot}/R_{min}$ arcsec (Robertson, Carter, \&
Dillinger 1991) as they pass the Sun and are observed at the earth, where $R_{min}$ 
is the  closest approach of the photon to the center of the Sun.  

     I approximately computed the (half) deflection as follows:  The 
orbit was computed
as a two-body problem with [7] using Butcher's 5th order Runge-Kutta method 
following Boulet (1991) with the constraint that $v = c$.  The path ran for
150 million km in the tangential direction starting from R$_{min}$ 700000 km. 
The time step was 0.005 s near the Sun and increased outward.  The Sun was 
modelled as a sphere with no internal motions, so no 
retardation.  The radial density distribution was interpolated from Lebretton 
\& Dappen (1988), normalized to the solar mass, and the force was integrated 
over 500 density shells, 500 latitudes, and 1000 longitudes.   
I integrated the force over the volume at each step and corrected the two-body 
force.  The deflection was found to be 1.751 $R_{\odot}/R_{min}$ arcseconds 
in agreement with observation.  Actually the same deflection is found by 
treating the Sun as a point mass, but the force is slightly weaker near the 
Sun, t $<$ 1.866 s, and slightly stronger further away.  Thus the deflection 
of photons is an effect of special relativity.

\section{Gravitational Redshift}

     The topic of this paper is the relativistic deflection of 
orbits.  One of the referees claimed that my treatment of the 
photon mass would not generate the observed gravitational redshift 
for the sun of 0.636 km/s because a photon moving radially away 
from a point mass experiences no force.  That is correct for a
point mass, for when $(\hat{c}\cdot\hat{r})^2 = 1, \ \vec{F} = 0$.
In Newtonian gravity a sphere with a radial density distribution 
produces the same force as a point mass at its center.  This is a
mathematical accident, not physics.  A photon at the surface of a 
sphere sees a whole $2\pi$ hemisphere, not a point.  Our experience 
with nearly spherical planets and stars has conditioned us to 
ignore the complexity in physics that cancels in spherical symmetry.
In addition, gravitational redshift is relative, $ \triangle \lambda / \lambda $.
It is independent of the mass (energy, freequency, wavelength) of the 
photon.  It is also independent of the ``shape" of the photon.

     A discussion of general relativity is given in Kurucz (2006).

\acknowledgments
     This work was supported, in part, by NASA grants NAG5-7014, NAG5-10864,
and NNG04GI75G.

\end{document}